\documentclass[a4paper,reprint,superscriptaddress,prb,aps]{revtex4-2}

\usepackage[english]{babel}
\usepackage[utf8]{inputenc}
\usepackage[T1]{fontenc}

\usepackage{amssymb}
\usepackage{amsmath}
\usepackage{amsthm}
\usepackage{graphicx}
\usepackage[hidelinks]{hyperref}
\usepackage{cleveref}
\usepackage[dvipsnames]{xcolor}
\usepackage{tikz}
\usetikzlibrary{shapes.geometric, arrows}
\usepackage[shortlabels]{enumitem}
\usepackage[normalem]{ulem}

\newcommand{\eps}{\varepsilon}
\newcommand{\vxc}{v_\mathrm{xc}}
\newcommand{\vxcbar}{\tilde{v}_\mathrm{xc}}
\newcommand{\rhoref}{\rho_\mathrm{gs}}

\newcommand{\DS}{\mathcal{D}}   %
\newcommand{\VS}{\mathcal{V}}   %

\newcommand{\abs}[1]{\left|#1\right|}
\newcommand{\norm}[1]{\left\Vert#1\right\Vert}
\newcommand{\dd}[2][\none]{\,\mathrm{d}^{#1}#2}

\DeclareMathOperator*{\argmin}{\mathrm{argmin}} 

\crefname{figure}{Fig.}{Figs.}
\Crefname{figure}{Figure}{Figures}

\crefname{equation}{Eq.}{Eqs.}
\Crefname{equation}{Equation}{Equations}

\begin{document}
\title{Kohn--Sham inversion with mathematical guarantees}

\author{Michael F. Herbst}
\email{michael.herbst@epfl.ch}
\affiliation{Mathematics for Materials Modelling (MatMat), Institute of Mathematics \& Institute of Materials,  École Polytechnique Fédérale de Lausanne, 1015 Lausanne, Switzerland}
\affiliation{National Centre for Computational Design and Discovery of Novel Materials (MARVEL), École Polytechnique Fédérale de
Lausanne, 1015 Lausanne, Switzerland}

\author{Vebjørn H. Bakkestuen}
\affiliation{Department of Computer Science, Oslo
Metropolitan University, 0130 Oslo, Norway}

\author{Andre Laestadius}
\email{andre.laestadius@oslomet.no}
\affiliation{Department of Computer Science, Oslo
Metropolitan University, 0130 Oslo, Norway}
\affiliation{Hylleraas Centre for Quantum Molecular Sciences, Department of Chemistry, University of Oslo, P.O. Box 1033 Blindern, N-0315 Oslo, Norway}

\begin{abstract}
We use an exact Moreau-Yosida regularized formulation 
to obtain the exchange-correlation potential for periodic systems. We
reveal a profound connection between rigorous mathematical principles and
efficient numerical implementation, which marks the first computation of a Moreau--Yosida-based inversion for physical systems.
We develop a mathematically rigorous inversion algorithm which is demonstrated for representative bulk materials,
specifically bulk silicon, gallium arsenide, and potassium chloride.
Our inversion algorithm allows the construction of rigorous
error bounds that we are able to verify numerically.
This unlocks a new pathway to analyze Kohn--Sham inversion methods,
which we expect in turn to foster mathematical approaches for developing
approximate functionals.
\end{abstract}

\maketitle

\section{Introduction}\label{sec:Intro}
Density-functional theory (DFT) simulations are an indispensable tool in chemistry,
materials science and solid-state physics~\cite{Burke2012,Verma_2020,Teale2022}.
In this theory, the unknown is the electronic density $\rho(\mathbf{r})$
instead of the full many-body wavefunction, making computations
substantially more tractable~\cite{Hohenberg1964}.
However, one of the key ingredients of DFT,
the universal density functional,
is not known explicitly~\cite{Verstraete2009}.
One therefore usually employs the Kohn--Sham (KS) formulation~\cite{KS1965},
where all unknowns of DFT are collected into the exchange-correlation~(xc) functional,
which is subsequently approximated.
Substantial work has been devoted to developing
accurate xc functional approximations~\cite{Toulouse2023}.
Although DFT is in principle exact, and  
despite significant advancements, KS-DFT still faces challenges in certain
physical contexts. Notable difficulties include accurately describing
processes involving fractional electronic charge,
such as dissociation or charge-transfer excitations~\cite{Cohen2008Science,Tempel2009,Helbig2009},
as well as addressing the
well-known band gap problem, where semiconductor band gaps are
underestimated~\cite{Sham1983PRL,Sham1985PRB,Perdew1982PRL,Perdew1983PRL,Gruning2006}.
Developing better xc functional approximations thus
remains a major research thrust~\cite{Mardirossian2017}.
One obstacle is a lack in mathematical understanding
between the exact universal density functional and common
approximations~\cite{Teale2022}, making the rigorous construction of new and better functionals hard.

In this work we will focus on KS inversion~\cite{Aryasetiawan1988,Knorr-Godby-1992,Gorling1992,ZP1993,Wang1993,ZMP1994,Knorr_1994,vanLeeuwen_1994,Yang_Wu_2002,Wu-Yang-2003,Peirs2003,Kadantsev2004,Bulat2007,Gaiduk2013,Wagner2014,JensenWasserman2017,Zhang2018,Ou2018,kumar2019universal,Kanungo2019,Kumar2020,Garrick2020,Callow2020,Shi2021,Erhard2022,Gould2023,Ravindran2024arXiv}, which has been suggested
as a tool to aid the construction of xc functionals~\cite{Shi2021,Kanungo2019,Gould2023}. 
In \citeyear{vanLeeuwen_1994}, \citet{vanLeeuwen_1994}
utilized an inversion scheme to improve approximations to the KS potential. More recently there has been considerable interest
in KS inversion, leading for example to
targeted software packages such as $n2v$~\cite{Shi2022} and \textit{KS-pies}~\cite{Nam2021}.
Both implement a variety of established inversion schemes,
beyond the van-Leeuwen--Baerends method
including the Zhao--Morrison--Parr~\cite{ZMP1994} (ZMP) and Wu--Yang~\cite{Wu-Yang-2003} methods.
While initial exploration of KS inversion has focused on isolated systems and molecules, 
recent works have begun to tackle a wide range of solid-state systems as well~\cite{Aouina_2023,Ravindran2024arXiv}.
In contrast to the standard KS formulation,
where one is equipped with an approximate xc potential  
and determines the density from a variational principle (forward problem),
KS inversion does the reverse:
given a ground-state density one seeks the exact xc potential $v_\mathrm{xc}$ which reproduces the density in an auxiliary non-interacting setting.
Unfortunately, KS inversion is far less studied than the forward KS-DFT problem.
Additionally, the development of a robust and efficient numerical scheme for KS inversion remains an open challenge~\cite{Shi2021,Nam2021,Shi2022,Crisostomo2023,Wrighton2023,Gould2023}.

In a recently established result~\cite{Penz_2023}, the xc
potential has been obtained as a mathematical limit
within the Moreau-Yosida (MY) regularized formulation of
DFT~\cite{Kvaal2014,KSpaper2018,Laestadius_2019_CDFT,KS_PRL_2019,PRLerrata}. 
This limit involves first finding the \emph{proximal density} (of a
ground-state density) and then utilizing the duality map between the density
and potential spaces.
This novel link is extremely promising as MY regularization deals with the
non-differentiability of the exact universal functional~\cite{Lammert2007}.
Notably, for both the aforementioned fractional electronic charge as well as the band-gap problems, 
a profound connection to the non-differentiability of KS-DFT
has been established in early works on DFT~\cite{Sham1983PRL,Sham1985PRB}.

Here, we focus on periodic systems, a setting that is particularly suited to obtain
an efficient numerical scheme for KS inversion 
that maintains the connection to a rigorous mathematical formulation.
Employing the Density-Functional ToolKit~\cite{DFTKpaper} (DFTK),
we are able to propose an implementation,
which is capable of performing KS inversion on systems of practical relevance,
such as bulk silicon, potassium chloride, and gallium arsenide.
Notably, in the context of realistic physical systems, this is
the first time a MY-based framework has been applied for KS inversion. 
Being equipped with a MY inversion procedure 
within a strict mathematical framework, we demonstrate how this 
inversion scheme opens up new insights into convergence properties and error analysis. 
In particular, we develop first rigorous error bounds for inverse KS problems. 
This paves the way for improved numerical analyses of KS inversion and
opens up novel opportunities for the development of density functionals.

The remainder of this article is structured as follows.
In \cref{subsec:MathBackground}, we briefly present the mathematical framework of the MY inversion scheme within a periodic setting. Furthermore, in \cref{subsec:Inversion} we relate the proximal density to the xc potential $v_\mathrm{xc}$ and
present an error analysis that includes rigorous bounds based on the proximal mapping. 
\Cref{sec:Numerics} discusses our numerical implementation
and results when applying the scheme to three bulk materials. Lastly, in \cref{sec:Conclusions} we give some concluding remarks.

\section{Mathematical Results}\label{sec:Theory}
\subsection{Background}\label{subsec:MathBackground}
Density-functional theory considers a system of $N$ electrons subjected to a scalar potential $v(\mathbf{r})$ with the corresponding (lifted) operator $\hat V = \sum_{j=1}^N v(\mathbf{r}_j)$. We let $\hat T $ and $\hat W$ denote the kinetic energy and two-body interaction operators, respectively. 
Let $E^\lambda(v)$ denote the ground-state energy for the Hamiltonian $\hat T + \lambda \hat W + \hat V$ and $F^\lambda(\rho)$ the \emph{exact}
universal functional, where
$\lambda$ is the coupling constant. For our purposes, it will be enough to consider $\lambda = 0$ and $\lambda = 1$. 
We will use $\langle v,\rho \rangle$ to denote the dual pairing between $\rho\in \DS$ and $v\in \VS =\DS^*$, where $\DS$ is a Banach space (complete normed space with norm $\Vert\cdot\Vert_{\DS}$) and $\DS^*$ its dual (i.e., the space of bounded, linear functionals on $\DS$ with induced norm $\Vert\cdot\Vert_{\VS}$).
In the Levy formulation~\cite{levy1979}, we can compute a constrained-search universal functional $\tilde{F}^\lambda$ from states that minimize all internal energy
contributions (i.e., coming from $\hat T + \lambda \hat W$) under a density constraint. 
In addition, the Lieb functional is defined as
\begin{equation} \label{eq:LTlambda}
    F^\lambda(\rho) = \sup_{v \in \VS} \{E^\lambda(v) - \langle v,\rho\rangle \}
\end{equation}
and will we take this formula as a definition of the universal functional in our setting. In general, $F^\lambda(\rho) \leq  \tilde{F}^\lambda (\rho)$ and by construction $F^\lambda$ is convex and lower semicontinuous. 
A set $S$ is convex if the line segment $t x + (1-t)y \in S$ whenever $x,y\in S$. A functional defined on a convex set $S$ is convex if for all $x,y\in S$ and $t \in (0,1)$ we have $f(t x + (1-t)y) \leq t f(x) + (1-t)f(y)$. Moreover, $f$ is lower semicontinuous if $f(x) \leq \liminf_{y\to x} f(y) $ whenever $y\to x$.

Now, $F(\rho) := F^1(\rho)$ is the exact universal functional of fully interacting electrons. In addition, at $\lambda=0$, 
$T (\rho) := F^0(\rho)$ is the kinetic-energy functional. Note that we follow the convention of omitting the superscript at $\lambda=1$ and using $T$ instead of the superscript $\lambda=0$. This style also applies to both $E$ and $\tilde F$. 
For our purposes, DFT can be viewed as an energy minimization problem: 
given an external potential $v$, find the corresponding density $\rho$ such that $E(v) = F(\rho) + \langle v,\rho \rangle$.   Employing a slightly different sign convention~\footnote{The reader familiar with the standard convex (Fenchel) conjugate $f^*(y) = \sup_{x}(\langle y,x\rangle - f(x))$, can then compare to the DFT-adapted definitions $E(v) = F^\wedge(v) = -F^*(-v)$ and $F(\rho) = E^\vee(\rho) = (-E)^*(-\rho)$.}, $E$ can be viewed as the Fenchel conjugate (or Legendre transform) of $F$, and $F$ as the conjugate of $E$. 
The ground-state density is then found by the
Hohenberg--Kohn \emph{variational principle}~\cite{Hohenberg1964} 
$$E(v) = \inf_{\rho\in \DS}\{ F(\rho) +\langle v,\rho \rangle  \}.$$ 
In particular, obtaining the ground-state density $\rho$ is equivalent to saturating the Fenchel--Young inequality $E(v) \leq F(\rho) + \langle v, \rho\rangle $.

Next, we introduce the \emph{Moreau--Yosida regularization}~\cite{Barbu-Precepanu} of a density functional $\mathcal F : \DS \to \mathbb{R}$ at $\rho_0\in \DS$ as the infimum of $\mathcal F$ and a penalty term,
\begin{equation}\label{eq:MYreg}
     \mathcal{F}^\varepsilon(\rho_0) = \inf_{\rho \in \DS} \left\{ \mathcal{F}(\rho) + \tfrac{1}{2\varepsilon} \norm{ \rho - \rho_0}_{\DS}^2\right\},
\end{equation}
where $\varepsilon> 0$ is the regularization parameter. We refer to \cref{eq:MYreg} as an \textit{infimal convolution}.
Our further discussion will be limited to the case when $\mathcal F$ is a convex and lower semicontinuous functional and $\DS$ a uniformly convex space (in fact a Hilbert space, but \emph{not} identified with its dual). 
If $\mathcal{F}=F$ is the exact universal functional, then the (exact) ground-state energy, $E(v)$, of some external potential, $v$, can still be computed from $ F^\eps$ using~\cite{KSpaper2018}
\begin{equation*}
    \begin{aligned}
        E^\eps(v) &= \inf_{\rho\in\DS}\left\{ F^\eps(\rho) + \langle v,\rho \rangle  \right\}, \\
        E(v) &= E^\eps(v) + \tfrac{\eps}{2} \norm{v}_{\VS}^2 .  
    \end{aligned}
\end{equation*}
The exact relation between $E(v)$ and $ E^\eps(v)$ follows from the fact that the MY regularization is an infimal convolution together with the underlying relation of Fenchel conjugate pairs. In this regard, regularizing $F$ with the MY transform is \emph{lossless} as far as the energy is concerned.

We consider systems which are 
periodic on a lattice with a unit cell denoted by $\Omega \subset \mathbb{R}^3$.
The function spaces of interest are the periodic Sobolev spaces~\cite[Section~2]{Cances_2012} $H_\mathrm{per}^s(\Omega,\mathbb{C})$, $s=\pm1$, equipped with the norm
\begin{equation}\label{eq:Hs-norm}
    \norm{u}_{H_\mathrm{per}^{s}}^2 = \sum_{\mathbf{G}} (1 + |\mathbf{G}|^{2})^s |\hat{u}_\mathbf{G}|^2,
\end{equation} 
where the sum is over the usual reciprocal lattice vectors $\mathbf{G}$. 
Henceforth, we assume that 
the densities $\rho $ are elements of $\DS = H_\mathrm{per}^{-1}$ such that 
the potentials $v$ are in $\VS =H_\mathrm{per}^1$. 
Denote by $J: \DS \to \VS$ the duality mapping, i.e., the canonical map from the primary space to its dual,
\begin{equation}\label{eq:Jmap}
J(\rho) = \left\{ v\in \VS : \Vert v \Vert_{\VS}^2 = \Vert \rho\Vert_{\DS}^2 = \langle v,\rho\rangle  \right\}.    
\end{equation}
By the definition of the duality mapping in \cref{eq:Jmap} and together with the particular choices of vector spaces $\DS$ and $\VS$ (with their norms given by Eq.~\eqref{eq:Hs-norm}), we obtain 
\begin{equation}\label{eq:Jfourier}
J(\rho) = \sum_{\mathbf{G}} \frac{\hat{\rho}_\mathbf{G} e_\mathbf{G}}{1 + \abs{\mathbf{G}}^2} 
\end{equation}
as the Fourier representation of $J(\rho)$. %
Here $\hat{\rho}_\mathbf{G}$ are the Fourier coefficients of $\rho$ and $e_\mathbf{G}$ the associated (normalized) basis vectors. It thus follows from a direct calculation that 
\begin{equation}\label{eq:Jrealsp}
J[\rho](\mathbf{r}) = (\Phi \ast \rho )(\mathbf{r}) = \int_{\mathbb{R}^3} \frac{\rho(\mathbf{r}')}{4\pi\abs{\mathbf{r} - \mathbf{r}'}} e^{-\abs{\mathbf{r} - \mathbf{r}'}}  \dd[3]{r'} ,
\end{equation}
where $\Phi(\mathbf{r}) = \exp{(-\abs{\mathbf{r}})}/ (4\pi \abs{\mathbf{r}})$ is the Yukawa potential. Thus, in the setting considered here, the duality mapping $J(\rho)$ is the convolution of the density $\rho$ with the Yukawa potential, which will be of importance when calculating the MY regularization.

\subsection{The Inversion Formula \& Error Bounds}\label{subsec:Inversion}
In the interest of an inversion algorithm, suppose we are given some accurate (and fixed)
ground-state density $\rho_\mathrm{gs}$ for some periodic system of interacting particles. 
Moreover, assume that this density is non-interacting $v$-representable, i.e., there exist a potential for which $\rho_\mathrm{gs}$ is also a non-interacting ground-state density. 
In practice, $\rho_\mathrm{gs}$ could be obtained from various sources, e.g., experimental
data, full configuration interaction~\cite{Tribedi_2023}, coupled-cluster~\cite{Wu-Yang-2003}, or quantum
Monte-Carlo~\cite{Knorr-Godby-1992,Knorr_1994,Aouina_2023} calculations. The goal is then to obtain the one-body potential of the
associated KS system, in which the aforementioned xc
potential $v_\mathrm{xc}$ is the critical unknown.

Recall the \emph{kinetic-only} functional $T(\rho)$ that was defined as the Legendre transform of the non-interacting ground-state energy $E^0(v)$, that is (with $\lambda=0$ in~\cref{eq:LTlambda})
\begin{equation}\label{eq:Tsup}
    T(\rho) = \sup_{v \in \VS} \left\{E^0(v) - \langle v,\rho\rangle \right\} .
\end{equation}
Then, together with the  Hartree contribution $E_\mathrm{H}(\rho)$ and the fixed external potential $v_\mathrm{ext}\in \VS$, the kinetic-only functional defines our convex and lower semicontinuous guiding density functional
\begin{equation}\label{eq:Fmodel}
   \mathcal{F}(\rho) = T(\rho) + E_\mathrm{H}(\rho) + \int_\Omega v_\mathrm{ext}\rho . 
\end{equation} 
As already remarked above, $T(\rho)$ is both convex and lower semicontinuous,  
whilst for $v_\mathrm{ext}\in \VS$, the linear (and therefore convex) map $\rho\mapsto \int_\Omega v_\mathrm{ext}\rho$ is even continuous. Note that the periodic Coulomb potential, defined as the solution to the (periodic) Poisson equation with a uniform background charge, is in $\VS$, which is also the case for the local potential term of typical pseudopotential approximations~\cite{Goedecker1996,Hartwigsen1998,VanSetten2018}.
Next, since
the periodic Hartree contribution 
$E_\mathrm{H}(\rho) = \sum_{\mathbf{G} \neq 0} \frac{\abs{\hat{\rho}_\mathbf{G}}^2}{\abs{\mathbf{G}}^2}$
is convex it remains to prove that it is lower semicontinuous.
To that end, suppose that $\Vert \rho^n - \rho \Vert_{H^{-1}_\mathrm{per}(\Omega)} \to 0$. 
Since strong convergence implies that the norms are convergent, we have
$\norm{\rho^n}_{H_\mathrm{per}^{-1}} \to \norm{\rho}_{H_\mathrm{per}^{-1}}$. 
However, $E_\mathrm{H}(\rho)$ can be bounded by $\norm{\rho}_{H^{-1}_\mathrm{per}}^2$ from both above and below 
using that $\abs{\mathbf{G}}^2\leq 1+\abs{\mathbf{G}}^2\leq \mathrm{constant}\times \abs{\mathbf{G}}^2$ for $\mathbf{G}\neq 0$, i.e., 
\begin{equation*}
     C_1 \norm{\rho}_{H_\mathrm{per}^{-1}}^2 \leq  E_\mathrm{H}(\rho) \leq C_2 \norm{\rho}_{H_\mathrm{per}^{-1}}^2 
\end{equation*}
for some constants $C_1,C_2>0$. 
This implies that $E_\mathrm{H}(\rho)$ is a continuous function in the $H^{-1}_\mathrm{per}(\Omega)$ topology, which directly implies lower semicontinuity.

From the above discussion, it is clear that our choice of $\mathcal F$ (\cref{eq:Fmodel}) is convex and lower semicontinous, and thus eligible to the Moreau--Yosida program outlined in \cref{subsec:MathBackground}. 
The key optimization problem of our inversion scheme is the minimization over 
$\rho\in  \DS$ of 
\begin{equation}\label{eq:ProxOpt}
    \mathcal{E}(\rho\,;\rho_\mathrm{gs}) = \mathcal{F}(\rho) + \tfrac{1}{2 \varepsilon} \norm{\rho-\rho_\mathrm{gs}}_{\mathcal{D}}^2.
\end{equation}
The functional $\rho \mapsto \mathcal{E}(\rho\,;\rho_\mathrm{gs})$
is strictly convex and lower semicontinuous on the Hilbert space $\DS$ and the minimum is attained at a unique point~\cite{Barbu-Precepanu} $\rho_\mathrm{gs}^\varepsilon = \argmin_\rho \mathcal{E}(\rho;\rho_\mathrm{gs})$,
referred to as the proximal density of $\rho_\mathrm{gs}$.
In the limit $\eps\to 0^+$, it holds from \cite[Prop.~1.146]{Barbu-Precepanu} that $\rho^\eps_\mathrm{gs} \to \rho_\mathrm{gs}$.
Since $\underline{\partial} \tfrac{1}{2}\norm{\rho}_{\DS}^2 = J(\rho)$,  we have the 
stationary condition
\[ \underline\partial \mathcal F(\rho^\eps_\mathrm{gs}) +\tfrac{1}{\eps} J(\rho^\eps_\mathrm{gs} - \rho_\mathrm{gs}) \ni 0\]
for the minimization of $\mathcal{E}$.
Here $\underline\partial \mathcal F$ denotes the subdifferential of $\mathcal F$ and is the collection of all tangent functionals of $\mathcal{F}$. 
Under the assumption that the zero element of $\VS$ belongs to the set $ \underline\partial \mathcal F(\rho_\mathrm{gs}) + \vxc$ for some $v_\mathrm{xc}$ (i.e., the assumption of non-interacting $v$-representability), we obtain $v_\mathrm{xc}$ as the limit of 
$\vxc^\eps = \tfrac{1}{\eps} J(\rho^\eps_\mathrm{gs} - \rho_\mathrm{gs})$ as \mbox{$\eps\to0^+$} by \cite[Thm.~2]{Penz_2023}. 
For our choice of function spaces, this implies using \cref{eq:Jrealsp} that
\begin{equation}\label{eq:vxc}
    v_\mathrm{xc}(\mathbf{r}) = \lim_{\varepsilon \to 0^+} \frac{1}{\varepsilon} \int_{\mathbb{R}^3} \frac{\rho^\varepsilon_\mathrm{gs}(\mathbf{r}')- \rho_\mathrm{gs}(\mathbf{r}')}{4\pi|\mathbf{r} - \mathbf{r}'|} e^{-|\mathbf{r} - \mathbf{r}'|} \dd[3]{r'}.
\end{equation}

The proximal mapping 
$\rho\mapsto \rho^\eps$ is a (firmly~\cite{FirmNonExpansive}) non-expansive operator.
To demonstrate this, we slightly generalize the proof of \cite[Prop.~2.3]{Barbu_2010} 
to the case considered here, i.e., Hilbert spaces not identified with their duals. 
From the general fact that $ -\tfrac{1}{\eps} J(\rho^\eps - \rho)$ is an element of $\underline\partial \mathcal F(\rho^\eps)$,
it follows that
\[ \varepsilon \left( \underline{\partial} \mathcal{F}^\varepsilon(\rho^\varepsilon) - \underline{\partial} \mathcal{F}^\varepsilon(\tilde{\rho}^\varepsilon) \right) \ni J(\rho - \rho^\varepsilon) - J(\tilde{\rho} - \tilde{\rho}^\varepsilon). \]
Then by the linearity of $J$, we have
$$\varepsilon \left( \underline{\partial} \mathcal{F}^\varepsilon(\rho^\varepsilon) - \underline{\partial} \mathcal{F}^\varepsilon(\tilde{\rho}^\varepsilon) \right) + J(\rho^\varepsilon - \tilde{\rho}^\varepsilon ) \ni J( \rho - \tilde{\rho}).$$  Then taking the dual pairing with $\rho^\varepsilon - \tilde{\rho}^\varepsilon$ yields   
\begin{equation} \label{eq:FirmNonExpans}
    \norm{\rho^\varepsilon - \tilde{\rho}^\varepsilon}^2_{\DS} \leq \langle J(\rho - \tilde{\rho}) , \rho^{\varepsilon} - \tilde{\rho}^\varepsilon\rangle,
\end{equation}
where we have used the  maximal monotonicity~\footnote{A set $A \subset X \times X^\ast$ is said to be monotone if 
\[\langle y_1 - y_2, x_1 - x_2\rangle \geq 0 \quad \forall [x_i,y_i] \in A, \, i = 1,2, \]
and maximally monotone if the set is not properly contained in any other monotone subset of $X \times X^\ast$~\cite[Def.~2.1]{Barbu_2010}. By the lower semicontinuity of the subdifferential of a convex functional, the maximal monotonicity follows.} 
of the subdifferential of $\mathcal{F}^\varepsilon$ and that $\langle J(\rho),\rho \rangle = \norm{\rho}^2_{\DS}$.
Using Hölder's inequality, we obtain
\begin{equation} \label{eq:ProximalIneq}
    \norm{\rho^\varepsilon - \tilde{\rho}^\varepsilon}_{\DS} \leq  \norm{\rho - \tilde{\rho}}_{\DS},
\end{equation}
which establishes the non-expansiveness of the proximal mapping. In fact, \cref{eq:FirmNonExpans} proves the \textit{firm} non-expansiveness~\cite{FirmNonExpansive}.

\Cref{eq:ProximalIneq} gives an estimate at a fixed \mbox{$\eps>0$}
for the error when performing the inversion
using  a perturbed density 
$ \tilde{\rho}_\mathrm{gs} = \rho_\mathrm{gs} + \Delta \rho$ rather than the exact reference $\rhoref $. 
For the purpose of further investigations, let us define the ratio
\[Q_\eps(\Delta\rho):= \frac{\Vert \rho_\mathrm{gs}^\eps - \tilde{\rho}_\mathrm{gs}^\eps \Vert_{\DS}}{\Vert  \Delta\rho \Vert_{\DS}}\leq 1.\] 
The ratio satisfies $Q_\varepsilon \to 1$ as $\varepsilon \to 0^+$, by the fact that $\rho^\varepsilon \to \rho$. 
The total error introduced in $v_\mathrm{xc}$ by using a $\tilde{\rho}_\mathrm{gs}$ and terminating the inversion at an $\varepsilon >0$ is by the triangle inequality
\begin{equation}\label{eq:TotalError}
    \norm{\vxc - \tilde{v}_\mathrm{xc}^\varepsilon }_{\VS} \leq \norm{v_\mathrm{xc}- v_\mathrm{xc}^\varepsilon}_{\VS} + \norm{v_\mathrm{xc}^\varepsilon - \tilde{v}_\mathrm{xc}^\varepsilon}_{\VS}. 
\end{equation}
The first term is the error arising from terminating the inversion at $\varepsilon>0$. This term is guaranteed to vanish as $\varepsilon \to 0^+$ by \cite[Thm.~2]{Penz_2023} and will be negligible for sufficiently small $\varepsilon$, a behavior also seen in our numerical calculations. Of interest in this article is the second term in \cref{eq:TotalError}, i.e., the error arising from the use of an inexact density. In the literature, errors in the energy caused by the use of perturbed densities are referred to as \textit{density-driven errors}~\cite{Kim_2013,Nam2020,Kaplan_2023}. We will therefore borrow the term ``density-driven'' when we speak of errors in the potentials caused by inexact densities.
Using that $v_\mathrm{xc}^\eps= \tfrac 1 \eps J(\rho_\mathrm{gs}^\eps - \rho_\mathrm{gs})$ and similarly for $\tilde{v}_\mathrm{xc}^\eps$, the error bound,
\begin{equation}\label{eq:Rinequality}
    \Vert \vxc^\eps - \vxcbar^\eps  \Vert_{\VS} \leq \frac{1 + Q_\eps(\Delta\rho)}{\varepsilon}\Vert \Delta \rho \Vert_{\DS},
\end{equation}
follows from the linearity of $J$, the triangle inequality, and \cref{eq:ProximalIneq}.
We remark that \cref{eq:Rinequality} is our main theoretical result and will be the focus of much of the numerical study in \cref{subsec:ErrorEstimates}. In particular, we will demonstrate numerically, that the ratio $Q_\varepsilon$ --- central to the above bound --- is indeed confined to $0 \leq Q_\varepsilon \leq 1$ for physical systems of interest.

We next discuss the implementation of the presented inversion scheme
as a practical inversion algorithm and verify the derived error bound, \cref{eq:Rinequality},
for a few representative solid-state systems.

\section{Numerical Results}\label{sec:Numerics}
\subsection{Implementation of the Inversion Algorithm}
In our inversion scheme the
potential $v_\text{xc}$ corresponding to the given density $\rho_\text{gs}$
is obtained by taking the $\varepsilon\to0^+$ limit in \cref{eq:vxc}.
To numerically extrapolate this limit we simply employ
an exponentially decreasing sequence in $\varepsilon$,
ranging between $1$ and about $10^{-7}$
and defer the investigation of a more tailored $\varepsilon$ sequences to future work.
For each $\varepsilon$ we minimize \cref{eq:ProxOpt} with respect to $\rho$
as will be detailed below. This yields the proximal density $\rho_\text{gs}^\varepsilon$
from which we extract $v_\mathrm{xc}^\varepsilon
= \frac1\varepsilon J(\rho_\text{gs}^\varepsilon - \rho_\text{gs})$,
i.e. by applying the duality mapping.
As $\varepsilon$ decreases, the
expectation is that $v_\mathrm{xc}^\varepsilon$ converges numerically to $v_\mathrm{xc}$.

The uniqueness of the proximal density for a fixed $\varepsilon>0$ implies that
$\rho^\varepsilon_\mathrm{gs}$ can be obtained from $\mathcal{E}$
of \cref{eq:ProxOpt}
by any minimization procedure of choice.
In agreement with the usual KS-DFT implementation for this setting
we will employ a parametrization of the proximal density in terms of
orthonormal orbitals $\Phi = (\psi_1, \ldots, \psi_{N_b})$,
i.e.~$\int_\Omega \psi_i^\ast \psi_j = \delta_{ij}$.
In this notation we do not make the $k$-points explicit,
i.e.~the sum over $i$ runs over all orbitals
at all reducible $\textbf{k}$-points used to discretize the Brillouin zone.
Moreover in this work we will only consider periodic insulators,
such that we can avoid making spin explicit and simply take
$N_b$ to be the product of the number of electron pairs and the number of $k$-points.
In this parametrization the density is thus obtained as
\[
    \rho_\Phi(\mathbf{r}) = 2 \sum_{i=1}^{N_b} |\psi_i(\mathbf{r})|^2
\]
enabling to rewrite \cref{eq:ProxOpt} as the minimization of
\begin{equation}\label{eq:Epractical}
\begin{aligned}
    \mathcal{E}(\Phi, \rho_\mathrm{gs})
    =&\, \sum_{i=1}^{N_b} \int_\Omega \left| \nabla \psi_i\right|^2  + E_\mathrm{H}(\rho_\Phi) \\
    &+ \int_\Omega v_\mathrm{ext} \rho_\Phi + \frac{1}{2\varepsilon} \| \rho_\Phi - \rho_\mathrm{gs} \|_{\mathcal{D}}^2
\end{aligned}
\end{equation}
with respect to the orbitals $\Phi$.
Notice that in contrast to \cref{eq:ProxOpt},
for the numerical implementation we use the standard KS kinetic energy expression in terms of orbitals
instead of the definition based on the Legendre transform discussed
in \cref{eq:Tsup}.

\Cref{eq:Epractical} has a similar structure
to the usual energy expression in KS-DFT. The only difference
is that the usual approximate xc functional is replaced
by a penalty term $\frac{1}{2\varepsilon} \| \rho_\Phi - \rho_\mathrm{gs} \|_{\mathcal{D}}^2$.
As a result standard direct minimization techniques
in KS-DFT~\cite{Payne1992,Kresse1996,Gonze1997,Mostofi2003,Vecharynski2015,Dai2017}
can be directly applied.
Here, we minimize $\mathcal{E}$ using a BFGS-based quasi-Newton scheme~\cite{Edelman1998,Boumal2023}
adapted to the geometrical structure (Stiefel manifold)
due to orthogonality constraint between the orbitals.
The optimization is stopped either if the optimizer
has been obtained to machine precision or if the change
in $\rho^\varepsilon_\mathrm{gs}$
between two iterations drops below $0.01 \varepsilon$.
Here, this heuristic criterion yields a good compromise
between the required computational time and the
accuracy of $v_\mathrm{xc}^\varepsilon$
as well as an excellent agreement with our theoretical error bounds
as will be shown in Section III D.
We defer an investigation of a more efficient
stopping criterion to future work.
Further details on the numerical procedure can be found on GitHub~\cite{FootnoteGitHub},
where the source code to reproduce the numerical results and all figures are available.

\subsection{Computational Setup}
\begin{figure}[tb]
    \centering
    \includegraphics[scale=1]{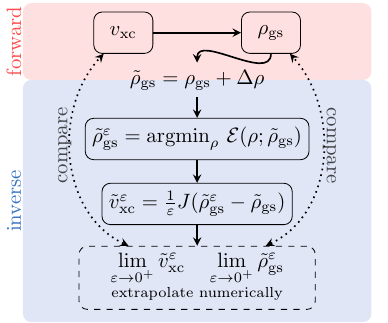}
    \vspace{-2ex}
    \caption{
        An illustration of the Moreau--Yosida (MY) inversion scheme. Given $v_\mathrm{xc}$, a ground-state density $\rho_\mathrm{gs}$ is first computed in a forward, reference calculation. 
        Here, $\Delta\rho$ represent an introduced error, resulting in an inexact reference density $\tilde{\rho}_\mathrm{gs}$.
        For this $\tilde{\rho}_\mathrm{gs}$, the proximal point $\tilde{\rho}^\eps_\mathrm{gs}$ is found from the MY regularization.
        An $\eps$-dependent potential is then found by means of the duality mapping $J$, whereupon the potential is obtained by taking $\eps\to 0^+$.  The results of the inversion can then be compared with  the forward scheme.}
        \label{fig:flowchart}
    \vspace{-3ex}
\end{figure}

For obtaining our numerical results
we follow the procedure outlined in \cref{fig:flowchart}.
We first perform a forward KS computation (red panel)
yielding both an xc potential $v_\text{xc}$ as well as a corresponding
reference ground-state density $\rho_\text{gs}$.
We employ the PBE xc functional~\cite{Perdew1996} as well as
the ``standard'' PBE pseudodojo pseudopotentials~\cite{VanSetten2018}
with non-linear core corrections.
Furthermore we use a $k$-point spacing of at most $0.12 \text{\r{A}}^{-1}$
and a rather tight kinetic energy cutoff
of about twice the value recommended for our selected pseudopotentials.
This enables us to reliably study the effect of controlled perturbations $\Delta \rho$
to the reference density $\rho_\text{gs}$ and test our error bounds (see Section III.D).
Finally, the obtained density $\rho_\text{gs} + \Delta \rho$ is fed to the KS inversion
procedure described in the previous Section III.B (\cref{fig:flowchart}, blue panel).
We remark that in this step, to achieve consistency between the forward
and inverse problem, the same pseudopotential approximation has been used.
As a result an additional non-local potential term is added to \cref{eq:Epractical}
corresponding to the Kleiman--Bylander part of the pseudodojo pseudopotentials.
For further details we refer to the reference implementation on GitHub~\cite{FootnoteGitHub}.

We remark that this setup enables us to directly compare
the extrapolated proximal density and the extrapolated xc potential to the respective
quantities $\rho_\text{gs}$ and $v_\text{xc}$ obtained in the forward calculation.
In the literature such a setup,
where the same quantum-chemical model and discretization
basis is employed for both the inversion and the forward calculation,
is sometimes referred to as an \textit{inverse crime}~\cite{Jensen_2016,JensenWasserman2017,Shi2021}.
However, we remark that our main focus is to highlight the strict mathematical
results offered by the MY inversion scheme.
It is thus beneficial to be able to compare both density and potential
from the inversion procedure against the reference from the forward scheme.

\subsection{Exact inversion}
\begin{figure}[tb]
        \centering
        \includegraphics[width=\linewidth]{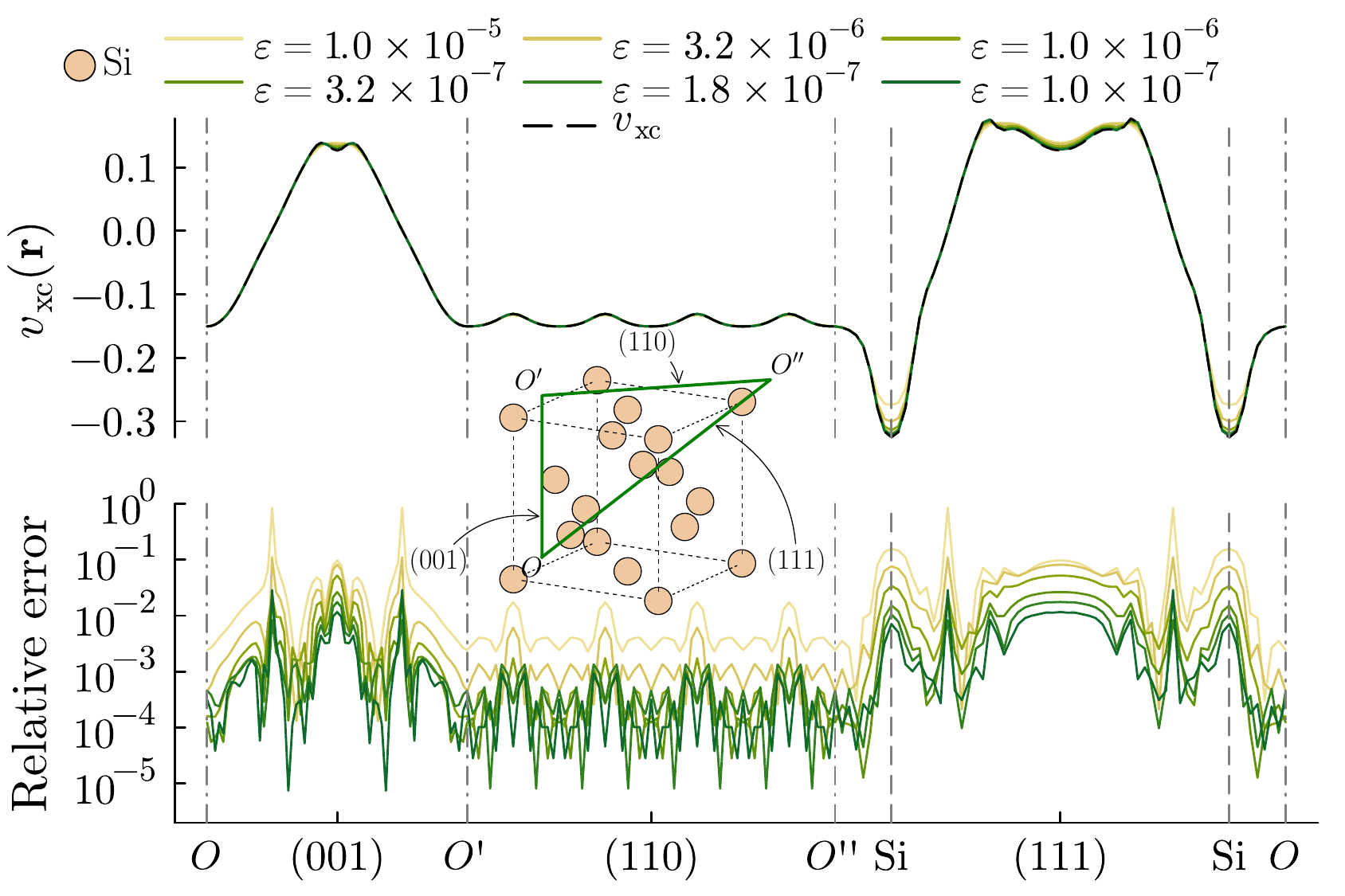}
        \vspace{-5ex}
        \caption{
        (top) Real-space plot of the reference xc potential along with the potential obtained from inversions for different values of the regularization parameter $\varepsilon$ for bulk silicon.  The potential is displayed along a closed path intersecting the high symmetry points~\cite{Chen_2021,FootnotePath}.
        (bottom) The associated pointwise relative error of the potentials obtained from inversions at various $\varepsilon$ compared to the reference xc potential. The crystalline structure and the path is shown in the inset.}
        \label{fig:vxc_silicon}
        \vspace{-3ex}
\end{figure}

\begin{figure}[tb]
        \centering
        \includegraphics[width=\linewidth]{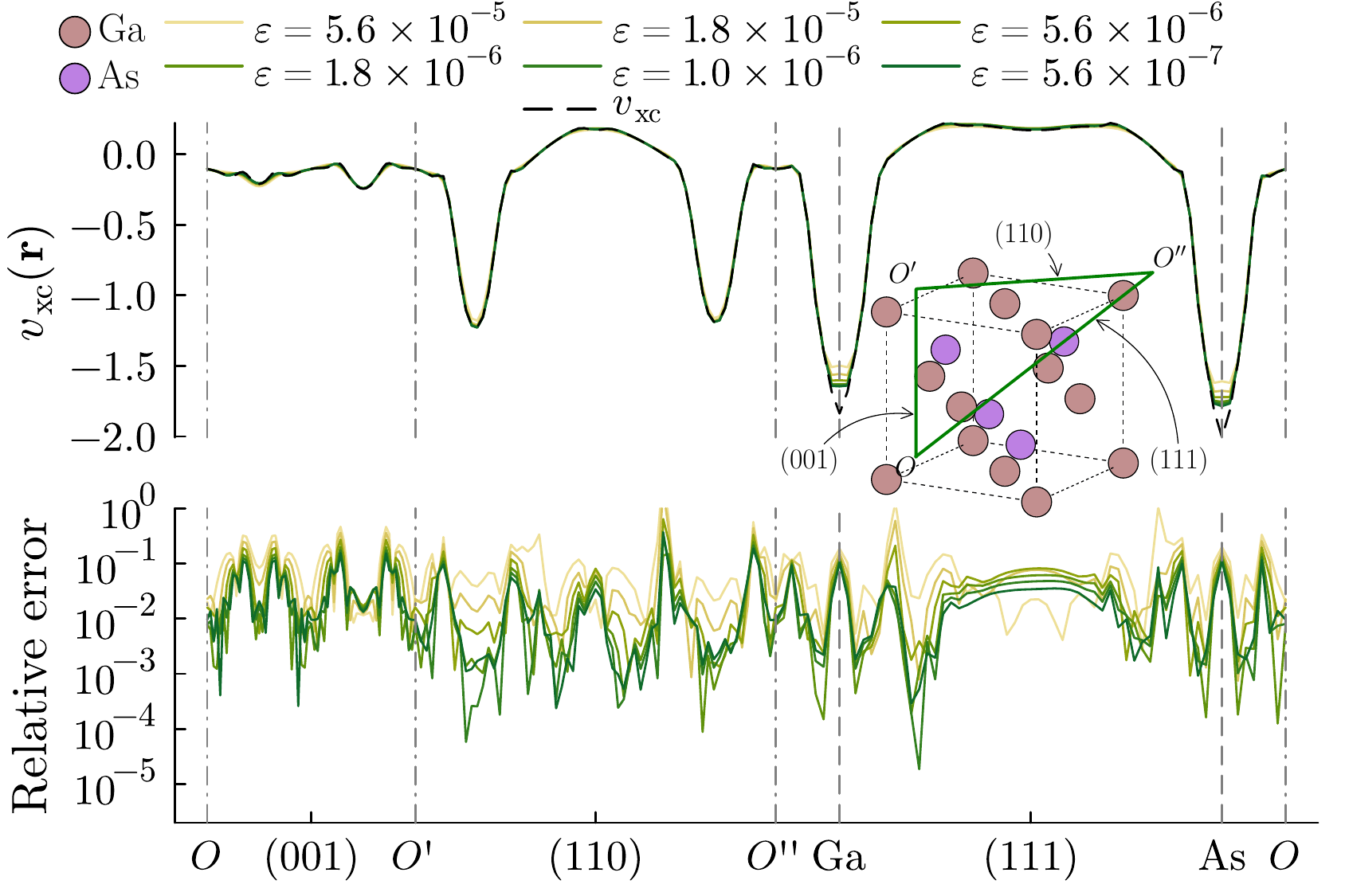}
        \vspace{-5ex}
        \caption{Analogous plots to \cref{fig:vxc_silicon} for gallium arsenide (GaAs) along the equivalent path shown in the inset crystalline structure.}
        \label{fig:vxc_gaas}
\end{figure}

\begin{figure}[tb]
        \centering
        \includegraphics[width=\linewidth]{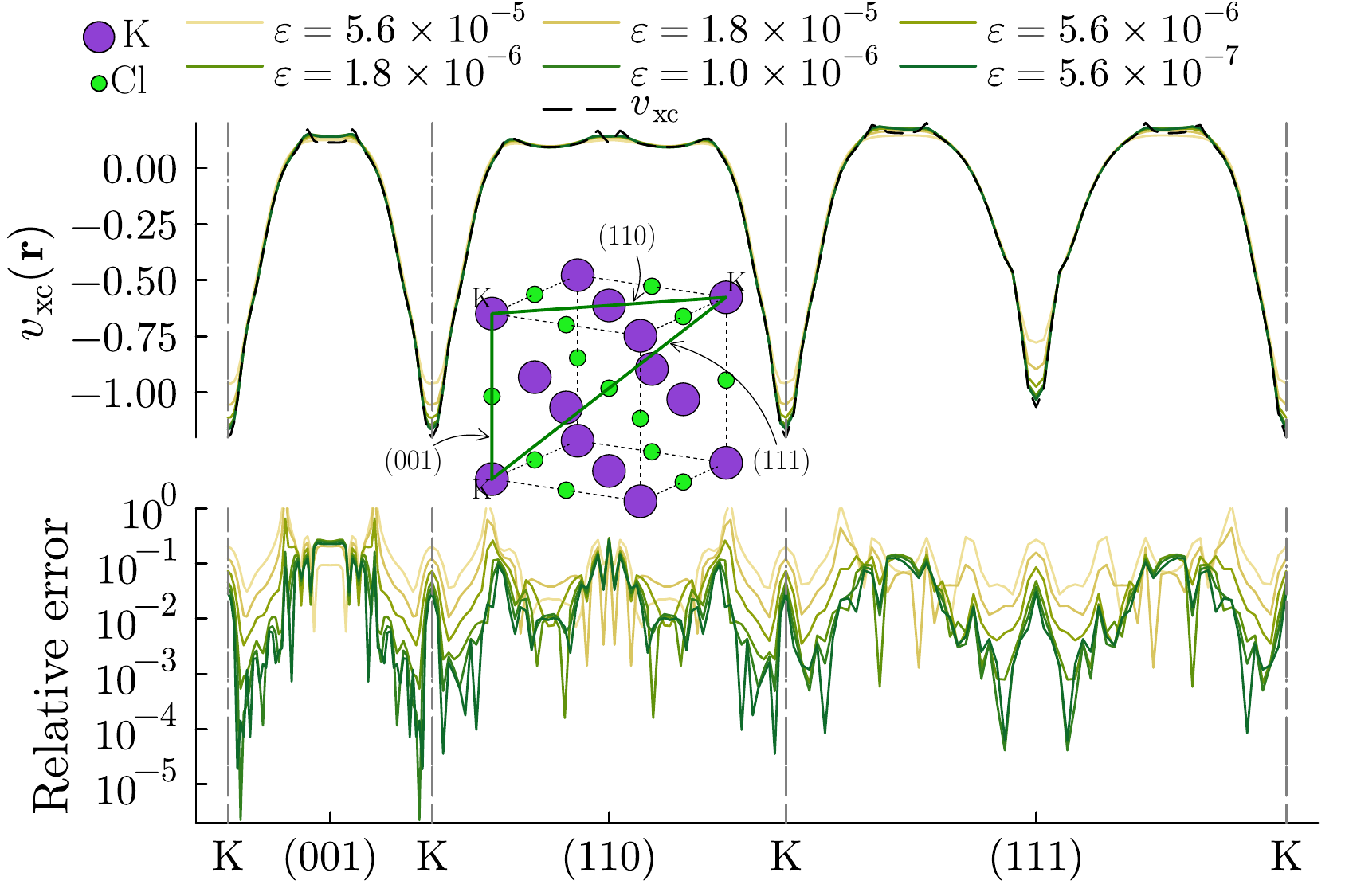}
        \vspace{-5ex}
        \caption{The equivalent plots to \cref{fig:vxc_silicon,fig:vxc_gaas} for potassium chloride (KCl) displayed along a similar path shown in the inset crystalline structure. }
        \label{fig:vxc_kcl}
\end{figure}

First we consider the case of an ``exact'' inversion:
we apply no additional noise ($\Delta \rho = 0$)
and expect our inversion scheme to fully recover the xc potential.
This setting we study on three bulk materials:
silicon (\cref{fig:vxc_silicon}) and
gallium arsenide (\cref{fig:vxc_gaas})
as well as potassium chloride (\cref{fig:vxc_kcl})
as an example for an ionic solid.

\Cref{fig:vxc_silicon} shows the resulting potentials for bulk silicon,
which are traced along the closed path in the unit cell
suggested by \citeauthor{Chen_2021}~\cite{Chen_2021,FootnotePath}.
The top panel contrasts the reference $v_\mathrm{xc}$ against
the potentials obtained from the inversion at selected values of
$\varepsilon$. The bottom panel shows the
pointwise absolute relative error of $v_\mathrm{xc}^\varepsilon$ against the
reference $v_\mathrm{xc}$. Two important features are observed.
(1) Our inversion procedure accurately recovers the potential numerically:
relative errors are below the 10 percentile for
$\varepsilon \sim 10^{-6}$, and decrease by another order of magnitude by
reducing $\varepsilon$ by an order of magnitude.
(2) Near the sharpest features of the potential,
the pointwise convergence in $\varepsilon$ is slower
and the relative errors larger.

\Cref{fig:vxc_gaas,fig:vxc_kcl} show the equivalent plots for gallium
arsenide (GaAs) and potassium chloride (KCl) demonstrating the applicability to
multiple standard material systems.
The xc potential for GaAs is plotted along the same path as for silicon,
starting between a Ga--Ga bond, while for KCl we start the trace
directly on a potassium (K) atom, see respective insets. 
Contrasting with silicon, the absolute
relative errors are slightly larger for GaAs and KCl at the same values of
$\varepsilon$. Still, overall the reference potentials are accurately recovered
(pointwise) across all three systems when no additional noise is applied
($\Delta= 0$). 

\subsection{Noisy inversion and error estimates}
\label{subsec:ErrorEstimates}
Having established the inversion algorithm to work well
for multiple material systems in the absence of noise, 
we study numerically the susceptibility of the inversion algorithm
to density-driven errors in the reference ground-state density, $\rho_\mathrm{gs}$.
Recall our notation
$\tilde{\rho}_\mathrm{gs} = \rho_\mathrm{gs} + \Delta \rho$ for an inexact density.
In practice, the erroneous part of the density, $\Delta \rho$,
may arise from various sources depending on the origin of $\rho_\mathrm{gs}$,
e.g., basis truncations, change of basis, insufficiently converged reference
densities, and experimental inaccuracies. Here, we limit ourselves
to perturbations arising from an interpolation of the density
utilizing a smaller plane-wave basis
and focus solely on the bulk silicon test case.

\begin{figure}[tb]
    \centering
    \includegraphics[width=1\linewidth]{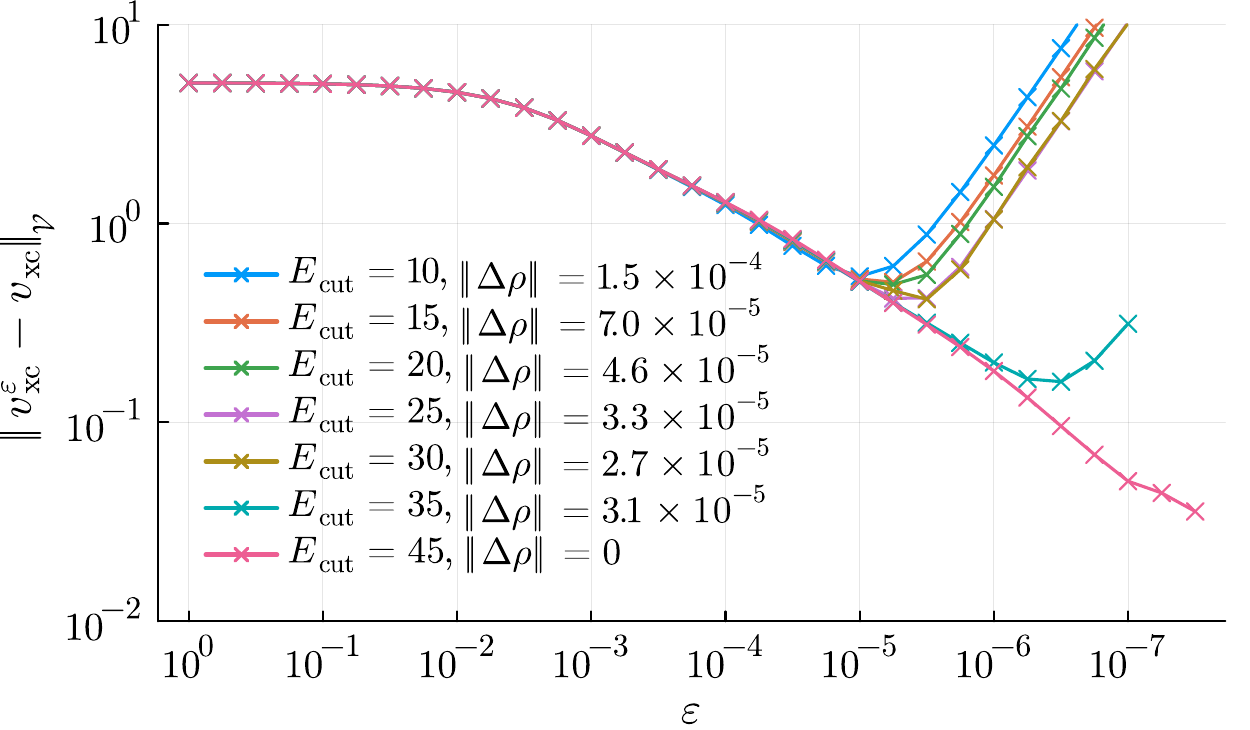}
    \vspace{-5ex}
    \caption{Convergence of $v_\mathrm{xc}^\varepsilon$ as a function of $\varepsilon$ for various perturbations introduced by basis truncations for bulk silicon. $E_\mathrm{cut}$ determines the cut off in the Fourier basis and $\norm{\Delta \rho}$ is the corresponding truncation error in the density in $\DS$-norm. $E_\mathrm{cut} = 45$ constitutes the (unperturbed) reference calculation.}
    \label{fig:ConvergenceTruncation}
    \vspace{-3ex}
\end{figure}

Our numerical calculations show that such perturbations to the reference
density do not alter the convergence properties of the potential
as long as $\varepsilon > \norm{\Delta \rho}_{L^2_\mathrm{per}}$,
see \cref{fig:ConvergenceTruncation}. In particular,
\cref{fig:ConvergenceTruncation} shows the $\VS$-norm difference of
$v_\mathrm{xc}^\varepsilon$ and $v_\mathrm{xc} $ for various truncations,
as indicated by the Fourier space cutoff
(essentially zeroing out the components for which
$\abs{\mathbf{G}}^2 > 2E_\mathrm{cut}$)
and the associated norm of the perturbation,
$\norm{\Delta \rho}_\DS$.
For smaller $\varepsilon$
the potential starts diverging from the reference in $\VS$-norm.

In order to further study the errors arising in the inversion scheme, recall our main error bound (\cref{eq:Rinequality}), 
$$ \Vert \vxc^\eps - \vxcbar^\eps  \Vert_{\VS} \leq \frac{1 + Q_\eps(\Delta\rho)}{\varepsilon}\Vert \Delta \rho \Vert_{\DS}.$$ 
We note that the right-hand side can be universally bounded by $\tfrac{2}{\eps}\Vert \Delta \rho \Vert_{\DS}$, i.e., the error in potential is small in the regime $\Vert \Delta\rho \Vert_{\DS} \ll \eps$.
To be able to further investigate $Q_\eps$, we also note that by the construction of $\vxc^\varepsilon$ (and $\vxcbar^\varepsilon$) and the linearity of $J$, it follows that
\begin{equation} \label{eq:Tinequality}
    \norm{ v_\mathrm{xc}^\varepsilon -  \tilde{v}_\mathrm{xc}^\varepsilon  - \tfrac{1}{\eps} J\left(\Delta\rho  \right)  }_{\VS}  \leq \frac{ Q_\varepsilon(\Delta \rho) }{\eps} \Vert \Delta \rho \Vert_{\DS}.
\end{equation}
Analogously to the ratio $Q_\varepsilon$, let us for \cref{eq:Rinequality,eq:Tinequality}, respectively, define the ratios 
\begin{align*}
    R_\varepsilon(\Delta \rho) &:= \varepsilon \frac{ \Vert \vxc^\eps - \vxcbar^\eps  \Vert_{\VS} }{\norm{\Delta \rho}_{\DS}},\\
    S_\varepsilon(\Delta \rho) &:= \varepsilon \frac{\norm{ v_\mathrm{xc}^\varepsilon -  \tilde{v}_\mathrm{xc}^\varepsilon  - \tfrac{1}{\eps} J\left(\Delta\rho  \right)  }_{\VS}}{\norm{\Delta \rho}_{\DS}}.
\end{align*}
Using that $\norm{J(\Delta\rho )}_{\VS}/\norm{\Delta\rho}_{\DS} =1$, the reverse triangle inequality applied to \cref{eq:Tinequality} gives $|R_\varepsilon-1|\leq Q_\varepsilon$, i.e,
\begin{equation} \label{eq:RBounds}
    0 \leq 1-Q_\varepsilon(\Delta \rho) \leq R_\varepsilon(\Delta \rho) \leq 1 + Q_\varepsilon(\Delta \rho) \leq 2.
\end{equation}  
Moreover, \cref{eq:Tinequality} implies that $0 \leq S_\varepsilon \leq Q_\varepsilon \leq 1$. 
Given the above, we remark that $Q_\varepsilon$ is a central quantity in our error analysis.

\begin{figure}[tb]
    \centering
    \includegraphics[width=1\linewidth]{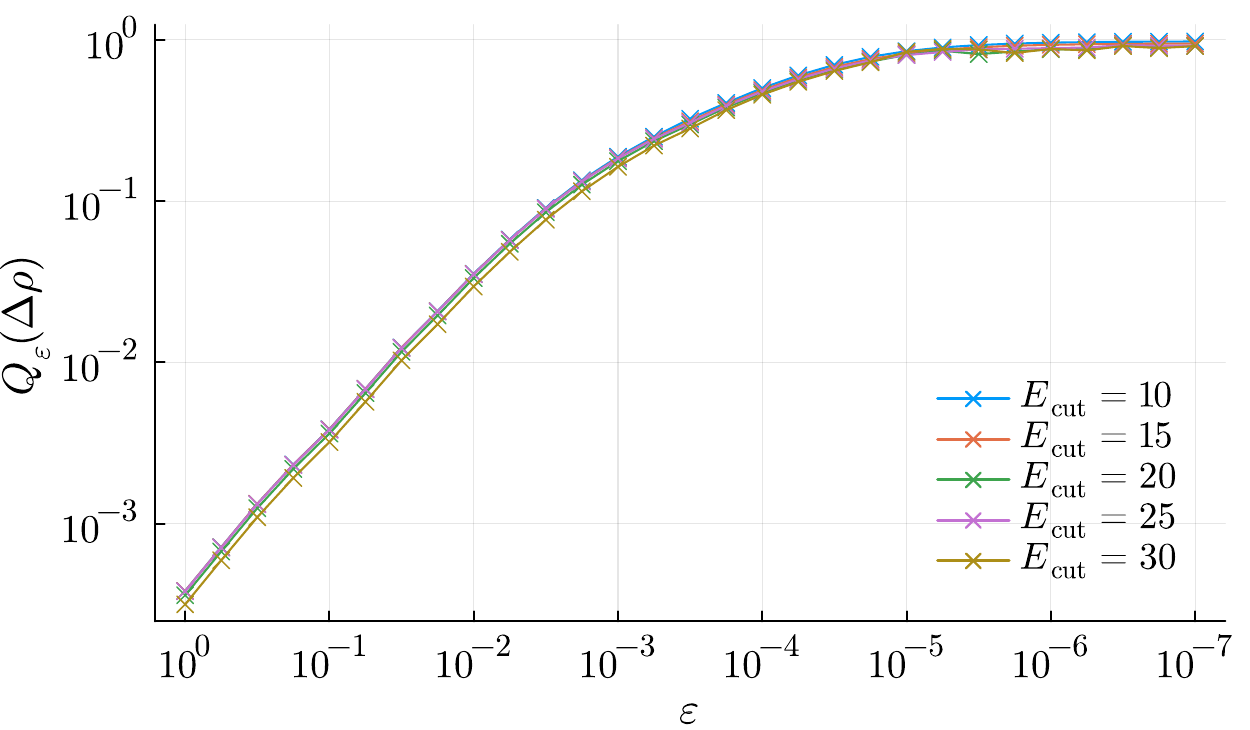}
    \vspace{-5ex}
    \caption{Related to the non-expansiveness of the proximal map, the ratio $ Q_\eps(\Delta\rho) = \norm{ \rho^\varepsilon_\mathrm{gs} - \tilde{\rho}^\varepsilon_\mathrm{gs} }_{\DS}/ \norm{\Delta \rho}_{\DS}$ is here shown for a decreasing sequence in $\varepsilon$. The corresponding values of $\norm{\Delta \rho}$ are shown in \cref{fig:ConvergenceTruncation}. The plot demonstrates the convergence of $Q_\varepsilon \to 1$ as $\varepsilon \to 0^+$ numerically for bulk silicon.}
    \label{fig:ContractionConstantTruncation}
    \vspace{-3ex}
\end{figure}

\begin{figure}[tb]
    \centering
    \includegraphics[width=1\linewidth]{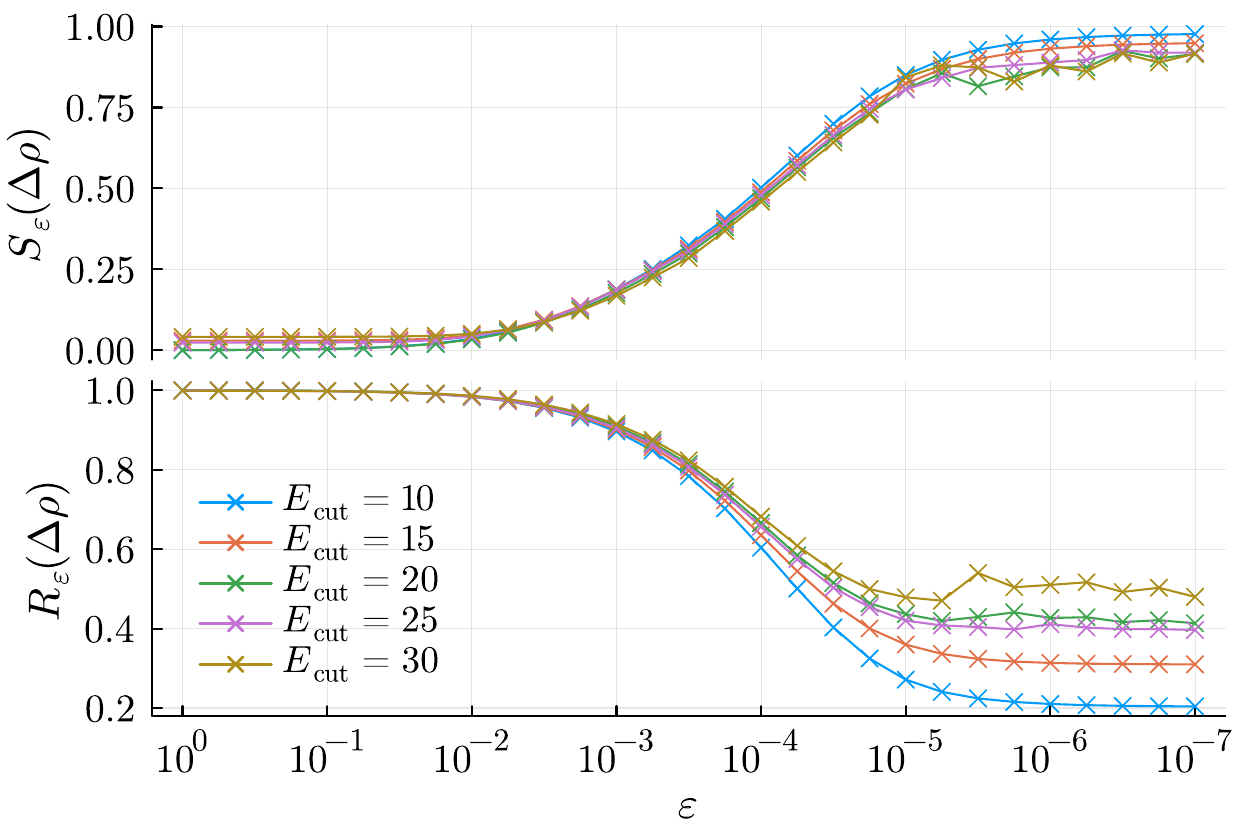} 
    \vspace{-5ex}
    \caption{
        The ratios  $S_\varepsilon$ and $R_\varepsilon$ as a function of
        $\varepsilon$ for various basis truncations on bulk silicon, whose $\norm{\Delta\rho}$
        are given in \cref{fig:ConvergenceTruncation}.
        (top) $S_\varepsilon$, a mathematical equivalent measure to $Q_\varepsilon$,  is small for large values of $\varepsilon$ and increases to one as $\varepsilon \to 0$.
        (bottom) For large $\varepsilon$-values $R_\varepsilon$ is similar to one, and approaches its lower bound (\cref{eq:RBounds})
        as $\varepsilon \to 0$.
        For $\varepsilon \lesssim 5 \times 10^{-6}$
        the problem becomes numerically challenging, which
        manifests in small oscillations in the trends
        of both quantities.}
        \label{fig:TruncationBounds}
        \vspace{-3ex}
\end{figure}

By comparing the proximal densities of $\rho_\mathrm{gs}$ and that of 
$\tilde{\rho}_\mathrm{gs}$, we can investigate the non-expansiveness of the proximal mapping by computing the ratio $Q_\varepsilon(\Delta\rho)$. 
A plot of $Q_\eps$ is shown in \cref{{fig:ContractionConstantTruncation}} for a series of perturbations induced by truncations of the density in the Fourier basis, where the corresponding size of the perturbation $ \norm{\Delta \rho}_{\DS}$ is shown in \cref{fig:ConvergenceTruncation}.  
By the proximal map, $Q_\varepsilon$ is bounded to the range $[0,1]$. From \cref{fig:ContractionConstantTruncation} we see that for large values of the regularization parameter ($\varepsilon \sim 1$)  $Q_\varepsilon \ll 1$, and $Q_\varepsilon \to 1^-$ as $\varepsilon \to 0^+$.
Additionally, we may obtain estimates for the error bounds on the xc potential, \cref{eq:Rinequality,eq:Tinequality}, directly from the ratios $R_\varepsilon$ and $S_\varepsilon$.   
In particular, \cref{fig:TruncationBounds} shows the ratios $R_\varepsilon$ and $S_\varepsilon$ (respectively bottom and top panels) obtained using the same $\tilde{\rho}_\mathrm{gs}$ as in \cref{fig:ContractionConstantTruncation}.  
Furthermore, \cref{fig:TruncationBounds} (top) shows that the ratio $S_\varepsilon$ depend on $\varepsilon$ in a very similar manner as $Q_\varepsilon$ (see  \cref{fig:ContractionConstantTruncation}). 
Note that for large values of $\varepsilon$ the ratio $S_\varepsilon$ does not adhere exactly to the bound $S_\varepsilon \leq Q_\varepsilon$ (\cref{eq:Tinequality}), as can be seen by comparing \cref{fig:ContractionConstantTruncation,fig:TruncationBounds}. However, this is not surprising given that it is obtained from the difference of three quantities that should pointwise be almost zero. 
For small values of $\varepsilon$, on the other hand, 
$S_\varepsilon$ is in excellent agreement with the bound set by $Q_\varepsilon$, also satisfying $S_\varepsilon \to 1$ as $\varepsilon \to 0^+$. 
Moreover, \cref{fig:TruncationBounds} (bottom) shows that the ratio $R_\varepsilon$ adheres strictly to the bound set by $Q_\varepsilon$ (\cref{eq:Rinequality}). In particular, $R_\varepsilon$ follow closely the lower bound $R_\varepsilon \geq 1-Q_\varepsilon $ (\cref{eq:RBounds}). 
Although obtaining the ratios
involved in the error bounds,
\cref{eq:Rinequality,eq:Tinequality,eq:RBounds}, is not computationally
feasible in all practical calculations, our numerical results conform
to the theoretical predictions.
In fact, \cref{fig:ContractionConstantTruncation,fig:TruncationBounds}
even indicate that the ratio $Q_\varepsilon$ may be estimated by a
constant independent of $\Delta \rho$, but parametrically dependent on
$\varepsilon$ (and the guiding functional $\mathcal{F}$).
Theoretically exploring this observation further 
marks a promising direction for future research towards more approximate bounds, which could be applied within a
practical inversion scheme to make it more efficient and reliable.
It is expected that improvements of the error bounds will enhance their applicability in practical calculations. Continued investigations will reveal if these first, but stringent, error estimates can be made useful in a broader sense.

\section{Conclusions}\label{sec:Conclusions}
In this article we set the stage for further in-depth
numerical analyses of KS inversion.
Facing the challenges of numerically solving the many-body
electronic-structure problem,
progress is often the result from combining insights
from mathematical analysis, more efficient numerical schemes
as well as physically sound approximations.
We have shown that the theoretical error bounds,
\cref{eq:Rinequality,eq:Tinequality,eq:RBounds}, as well as the
non-expansiveness, \cref{eq:ProximalIneq}, are manifested
in our numerical calculations.
In our numerical study we targeted three bulk materials systems:
two semiconductors (silicon and gallium arsenide)
as well as an ionic salt (potassium chloride),
demonstrating the applicability of our scheme to realistic systems.
Since the calculations have been performed in a general DFT code
with demonstrated performance on systems up to a few hundred electrons,
a generalization to more involved systems is possible and aspired for.
Future works should also aim to apply our presented inversion scheme
to densities obtained from accurate sources other than forward KS algorithms.
In this regard, considering reference densities from mean-field-like theories
beyond semi-local DFT similar to recent work~\cite{Ravindran2024arXiv}
represents a possible and interesting endeavour.
In our numerical approach we follow the standard practice of plane-wave
DFT to employ a pseudopotential for modelling the electron-nuclear interaction.
While this approach is a practical necessity for efficient calculations,
our current theoretical framework does not yet include
the effects of non-local potentials.
Closing this theoretical gap is an interesting direction for future research.
We would also like to briefly comment on our choice of function spaces
for the densities and potentials in our mathematical formulation. 
By choosing a potential space where the norm also measures derivatives, we naturally penalize oscillations in the targeted xc potential. 
In addition, the choice of function space also dictates the form of the duality mapping,
which in our setting leads to the numerically tractable form of a Yukawa kernel.
Other options are of course possible and future investigations may reveal better choices.

The inversion scheme, the explicit form of $v_\mathrm{xc}$ (\cref{eq:vxc}), the use of the 
non-expansiveness of the proximal map, and the error bounds on the inverted
potential are all novel theoretical results that are also numerically demonstrated.
These results establish mathematical guarantees for KS inversion not previously
seen, and also mark the first successful application of the MY framework to physical systems.
We believe that such mathematical results, closely accompanied with numerical
calculations, will significantly aid the development of more reliable KS inversion schemes.
Moreover, the first stringent error bounds presented here open a new avenue for
developing rigorous error estimates for density-potential inversion problems.
Strengthened KS inversion schemes, in union with error analysis, should
ultimately further the development of approximate density functionals.
Furthermore, as highlighted by \citeauthor{Shi2021}~\cite{Shi2021}, such
enhanced understanding of density-potential inversion may lead to advances in
explorations of the Hohenberg--Kohn mapping, quantum embedding techniques, and
enhanced optimised effective potentials.

~\newline
\emph{Data Availability:} Raw simulation data as well as source code to
reproduce all numerical results and the plots presented in this article are
openly available under DOI \href{https://doi.org/10.5281/zenodo.14894064}{10.5281/zenodo.14894064}
or the GitHub repository \texttt{mfherbst/supporting-my-inversion}~\cite{FootnoteGitHub}.

\begin{acknowledgments}
AL and VHB were supported by the ERC-2021-STG grant agreement No. 101041487 REGAL.
AL was also supported by the Research Council of Norway through
CoE Hylleraas Centre for Quantum Molecular Sciences Grant No. 262695 and CCerror Grant No. 287906.
MFH acknowledges support by the NCCR MARVEL, a National Centre of
Competence in Research, funded by the Swiss National Science
Foundation (Grant No. 205602).
MFH and AL thank the Oberwolfach Research Institute for Mathematics 
where part of this research was conducted. 
We express our gratitude for fruitful discussions with Nicola Marzari and Markus Penz.
\end{acknowledgments}

\bibliography{refs}

\begin{thebibliography}{83}
\bibitem [{Fir()}]{FirmNonExpansive}%
  \BibitemOpen
  \href@noop {} {}\bibinfo {note} {We use the notion of a firmly non-expansive
  operator as in the case of a Hilbert space identified with its dual with the
  obvious inclusion of the duality mapping, i.e.,
  \cref{eq:FirmNonExpans}.}\BibitemShut {Stop}%
\end{thebibliography}

\end{document}